\documentstyle[12pt]{article}
\newcommand\lesssim{\mathrel{\rlap{\lower4pt\hbox{\hskip1pt$\sim$}}
	\raise1pt\hbox{$<$}}}
\newcommand\gtrsim{\mathrel{\rlap{\lower4pt\hbox{\hskip1pt$\sim$}}
	\raise1pt\hbox{$>$}}}

\begin{document}

\title{Affleck-Dine Baryogenesis \\ after \\ Thermal Inflation}
\author{
E. D. Stewart \\
{\em Research Center for the Early Universe, University of Tokyo} \\
{\em Tokyo 113, Japan}
\and M. Kawasaki \\
{\em Institute for Cosmic Ray Research, University of Tokyo} \\
{\em Tanashi 188, Japan}
\and T. Yanagida \\
{\em Department of Physics, University of Tokyo} \\
{\em Tokyo 113, Japan} }
\maketitle
\begin{abstract}
We argue that an extension of the Minimal Supersymmetric Standard
Model that gives rise to viable thermal inflation, and so does not
suffer from a Polonyi/moduli problem, should contain right-handed
neutrinos which acquire their masses due to the vacuum expectation
value of the flaton that drives thermal inflation.
This strongly disfavours ${\rm SO}(10)$ Grand Unified Theories.
The $\mu$-term of the MSSM should also arise due to the vev of the
flaton.
With the extra assumption that $ m_L^2 - m_{H_u}^2 < 0 $, but of
course $ m_L^2 - m_{H_u}^2 + |\mu|^2 > 0 $, we show that a
complicated Affleck-Dine type of baryogenesis employing an $LH_u$
$D$-flat direction can naturally generate the baryon asymmetry of
the Universe.
\end{abstract}
%\pacs{98.80.Cq,11.30.Fs,12.60.Jv,14.60.St,14.80.Ly}
\vspace*{-113ex}
\hspace*{\fill}{RESCEU-8/96}\hspace*{3.55em}\\
\hspace*{\fill}{ICRR-Report-358-96-9}\\
\hspace*{\fill}{UT-744}\hspace*{6.7em}\\
\hspace*{\fill}{hep-ph/9603324}\hspace*{2.25em}
\thispagestyle{empty}
\setcounter{page}{0}
\newpage
\setcounter{page}{1}

\section{Introduction}

Thermal inflation \cite{ti,Ed,old} provides the most compelling
solution to the moduli (Polonyi) problem \cite{problem,Dine,recent}.
However, for a theory of the early Universe to be viable it must be
capable of producing a baryon asymmetry \cite{baryon}
\begin{equation}
\label{baryon}
\frac{n_B}{s} \sim 3 \times 10^{-11}
\end{equation}
by the time of nucleosynthesis.
Thermal inflation probably dilutes any pre-existing baryon asymmetry
to negligible amounts, and the final reheat temperature after thermal
inflation ($ T_{\rm f} \sim $ few GeV) is probably too low even for
electroweak baryogenesis.
Thus if thermal inflation really is the solution of the moduli
problem, then it is likely also to be responsible for baryogenesis.

In Section~\ref{ti} we explain why the flaton that gives rise to
thermal inflation probably also generates the masses of right-handed
neutrinos as well as the $\mu$-term of the Minimal Supersymmetric
Standard Model (MSSM). We also note the various ways in which a
potential domain wall problem can be avoided.
In Section~\ref{AD} we describe how a somewhat complicated
Affleck-Dine type mechanism can naturally generate the required
baryon asymmetry after thermal inflation.
In Section~\ref{con} we give our conclusions.

\section{Thermal inflation, right-handed neutrinos, and the
$\mu$-term}
\label{ti}

\subsection{Thermal inflation and right-handed neutrinos}

The superpotential of the MSSM is \cite{susy}
\footnote{
Here, and throughout most of this paper, all indices (the usual gauge
and generation indices, as well as any singlet indices) have been
suppressed. We will for the most part be focusing on the third
generation as is suggested by our notation.}
\begin{equation}
W_{\rm MSSM} = \lambda_t Q H_u t + \lambda_b Q H_d b
 + \lambda_\tau L H_d \tau + \mu_H H_u H_d
\end{equation}
Thermal inflation \cite{ti} requires that there is in addition at
least one flaton $\phi$ with vacuum expectation value $ |\phi| = M $
in the range
$ 10^{10} \,{\rm GeV} \lesssim M \lesssim 10^{12} \,{\rm GeV} $,
the lower bound coming from the requirement that thermal inflation
sufficiently dilutes the moduli, and the upper bound from the
requirement that the final reheat temperature after thermal inflation
$T_{\rm f}$ be high enough to thermalise the lightest supersymmetric
particles (LSP's) produced in the flaton's decay and so avoid an
excess of LSP's.

Also, in order for $\phi$ to be held sufficiently strongly\footnote{
Note the strong dependence on $T_{\rm C}$ in Eq.~(35) of
Ref.~\cite{ti}.}
at $\phi=0$ by the finite temperature during thermal inflation,
$\phi$ must have unsuppressed couplings to at least one other field,
say $\psi$, that is light when $\phi=0$.
We therefore either require a term $ \lambda_\phi \phi \psi^2 / 2 $
with $ \left| \lambda_\phi \right| \sim 1 $ in the superpotential,
or require $\phi$ to spontaneously break a continuous gauge symmetry
with gauge coupling $ g_\phi \sim 1 $
(with $\psi$ being the gauge field in this case).
One reason to prefer the Yukawa coupling over the gauge coupling is
that the renormalisation group \cite{susy} effect of the Yukawa
coupling would be to drive the soft supersymmetry breaking mass
squared of $\phi$ negative at low energies as is required for a
flaton, while the gauge coupling would have the opposite effect.
Another reason is that the gauge symmetry has no independent
motivation, while, as we shall see, the Yukawa coupling is very well
motivated. We will therefore focus on the case of the Yukawa
coupling.

After $\phi$ acquires its vacuum expectation value $M$,
$\psi$ will acquire a mass
$ \left| \lambda_\phi \right| M \sim 10^{10} \,{\rm GeV} $
to $ 10^{12} \,{\rm GeV} $ and so is not a MSSM field.
In order for $\phi$ to be coupled to the thermal bath, which is
presumably composed of MSSM fields, we therefore require $\psi$ to
couple to the MSSM.  
We therefore\footnote{
Assuming $\psi$ is not charged with respect to the MSSM continuous
gauge symmetries as this would, in general, destroy susy GUT gauge
coupling unification.
However, if $\psi$ is a complete ${\rm SU}(5)$ multiplet the
unification of the gauge couplings will be unaffected.
For example, one could have
$ \lambda_\phi \phi \psi^2 = \lambda_\phi \phi \bar{5} 5 $.
Alternatively, appropriate choices of representations could shift the
unification scale to the string scale \cite{Bachas}.}
require at least one\footnote{
If we have both then we need two different $\psi$'s, one charged and
one neutral under R-parity.}
of the terms $ L H_u \psi $ or $ \psi H_u H_d $ in the
superpotential, as these are the only possible renormalisable
couplings of a singlet to the MSSM.

The former is the standard coupling $ \lambda_\nu L H_u \nu $ of a
right-handed neutrino $ \psi = \nu $ to the MSSM, and furthermore
$\nu$ automatically acquires a mass
$ M_\nu = \left| \lambda_\phi \right| M \sim 10^{10} \,{\rm GeV} $ to
$ 10^{12} \,{\rm GeV} $ in the right range for the seesaw mechanism
\cite{seesaw} to generate a left-handed tau neutrino mass
\begin{eqnarray}
\label{seesaw}
m_{\nu_L}
& = & \frac{ m_{\rm D}^2 }{ M_\nu } =
\frac{ \left| \lambda_\nu \right|^2 \left( 174 \,{\rm GeV} \right)^2
\sin^2 \beta }{ \left| \lambda_\phi \right| M } \\
\label{mnu}
& \sim & 5 \,{\rm eV}
\left( \frac{ 3 \times 10^{12} \,{\rm GeV} }
{ M / \left| \lambda_\nu \right|^2 } \right)
\left( \frac{ 1 }{ \left| \lambda_\phi \right| } \right)
\left( \frac{ \sin^2 \beta }{ 0.5 } \right)
\end{eqnarray}
suitable for the mixed dark matter scenario for the formation of the
large scale structure of the Universe \cite{mdm}.

The latter coupling $ \psi H_u H_d $ does not have any good
independent motivation, apart from the fact that it is possible.
We will therefore focus on the former case. Later though, we shall
make use of this possibility for coupling to the MSSM in a different
context.

\subsection{The final reheat temperature after thermal inflation and
the $\mu$-term}

Now that we have a more precise picture of the couplings of the
flaton,
\begin{equation}
W_{\mbox{\scriptsize so far}} = W_{\rm MSSM} + \lambda_\nu L H_u \nu
 + \frac{1}{2} \lambda_\phi \phi \nu^2
\end{equation}
we can hope to make a more precise estimate of the decay rate of the
flaton and so the final reheat temperature after thermal inflation
$T_{\rm f}$.

We first note that the effective superpotential coupling
\begin{equation}
\label{wseesaw}
W_{\rm seesaw} =
 - \frac{ \left( \lambda_\nu L H_u \right)^2 }{ 2 \lambda_\phi \phi } 
\end{equation}
obtained by integrating out $\nu$, \mbox{i.e.} eliminating $\nu$ via
the constraint $ \partial W / \partial \nu = 0 $, will give a decay
rate of order $ \Gamma \sim m^5 / M^4 $ which is negligible.

The $\phi$ dependence of the low energy renormalised coupling
constants will give larger decay rates.
To estimate these we first need to know the contributions of the
right-handed neutrinos to the renormalisation group equations.
They are \cite{rg}
\begin{eqnarray}
16 \pi^2 \frac{d}{dt} \lambda_t
& = & \left| \lambda_\nu \right|^2 \lambda_t + \ldots \\
16 \pi^2 \frac{d}{dt} \lambda_\tau
& = & \left| \lambda_\nu \right|^2 \lambda_\tau + \ldots \\
16 \pi^2 \frac{d}{dt} \mu_H
& = & \left| \lambda_\nu \right|^2 \mu_H + \ldots \\
16 \pi^2 \frac{d}{dt} m_L^2
& = & 2 \left| \lambda_\nu \right|^2 \left( m_L^2 + m_\nu^2
+ m_{H_u}^2 + \left| A_{LH_u\nu} \right|^2 \right) + \ldots \\
16 \pi^2 \frac{d}{dt} m_{H_u}^2
& = & 2 \left| \lambda_\nu \right|^2 \left( m_L^2 + m_\nu^2
+ m_{H_u}^2 + \left| A_{LH_u\nu} \right|^2 \right) + \ldots \\
16 \pi^2 \frac{d}{dt} A_{QH_ut}
& = & 2 \left| \lambda_\nu \right|^2 A_{LH_u\nu} + \ldots \\
16 \pi^2 \frac{d}{dt} A_{LH_d\tau}
& = & 2 \left| \lambda_\nu \right|^2 A_{LH_u\nu} + \ldots \\
16 \pi^2 \frac{d}{dt} A_{\mu_HH_uH_d}
& = & 2 \left| \lambda_\nu \right|^2 A_{LH_u\nu} + \ldots
\end{eqnarray}
where $t$ is the logarithm of the renormalistion scale,
the $m$'s are the soft supersymmetry breaking masses,
the $A$'s are the soft supersymmetry breaking parameters in the terms
in the scalar potential of the form $ A W + \mbox{c.c.} $,
the $m$'s and the magnitudes of the $A$'s are of the order of the
soft supersymmetry breaking scale $ m_{\rm s} \sim 10^2 $ to
$ 10^3 \,{\rm GeV} $, and the \ldots\ stand for other terms
independent of the right-handed neutrinos.
$|\phi|$ sets the threshold for the right-handed neutrinos, and so
writing $ |\phi| = M + \delta\phi_r / \sqrt{2} $ we get the effective
couplings
\begin{equation}
W_{\rm eff} =
\frac{ \left| \lambda_\nu \right|^2 }{ 16 \sqrt{2}\, \pi^2 M }
\left( \lambda_t Q H_u t + \lambda_\tau L H_d \tau + \mu_H H_u H_d
\right) \delta\phi_r
\end{equation}
and
\begin{eqnarray}
V_{\mbox{\scriptsize soft eff}} & = &
\frac{ \left| \lambda_\nu \right|^2 }{ 8 \sqrt{2}\, \pi^2 M }
\left( m_L^2 + m_\nu^2 + m_{H_u}^2 + \left| A_{LH_u\nu} \right|^2
\right)
\left( \left| L \right|^2 + \left| H_u \right|^2 \right)
\delta\phi_r
\nonumber \\ && \mbox{}
+ \frac{ \left| \lambda_\nu \right|^2 }{ 8 \sqrt{2}\, \pi^2 M }
A_{LH_u\nu}
\left( \lambda_t Q H_u t + \lambda_\tau L H_d \tau + \mu_H H_u H_d
\right) \delta\phi_r
\end{eqnarray}
From these couplings we estimate the total decay rate to be
\begin{equation}
\Gamma \sim
\frac{ \left| \lambda_\nu \right|^4 m_{\rm s}^3 }{ 10^4 M^2 }
\end{equation}
This would give a final reheat temperature after thermal inflation of
\begin{eqnarray}
T_{\rm f} & \simeq &
g_*^{-\frac{1}{4}} \Gamma^{\frac{1}{2}} M_{\rm Pl}^{\frac{1}{2}} \\
& \sim & 10\,{\rm MeV}
\left( \frac{ 3 \times 10^{12}\,{\rm GeV} }
{ M / \left| \lambda_\nu \right|^2 } \right)
\left( \frac{ m_{\rm s} }{ 300\,{\rm GeV} } \right)^{\frac{3}{2}}
\end{eqnarray}
where the first bracket is constrained to be of order one, or perhaps 
less, by Eq.~(\ref{mnu}).
However, in order not to over-produce LSP's we require \cite{LSP}
\begin{equation}
T_{\rm f} \gtrsim 1\,{\rm GeV} \left(
\frac{ m_{\rm s} }{ 300\,{\rm GeV} } \right)^{1.5 \;{\rm to}\; 2}
\end{equation} 

Thus our model seems to be in trouble unless we can add some extra
coupling that gives a stronger decay rate.
The only possibility is to couple $\phi$ to $ H_u H_d $ in the
superpotential.
A term $ \phi H_u H_d $ would require a very small coupling constant
to avoid generating too large a $\mu$-term, but a term\footnote{
A term $ \lambda_\mu \phi^3 H_u H_d / M_{\rm GUT}^2 $ would be an
alternative if $ M \gtrsim 3 \times 10^{11}\,{\rm GeV} $.
We assume the displayed case for simplicity.}
\begin{equation}
\label{wdecay}
W_{\rm decay} = \frac{\lambda_\mu \phi^2 H_u H_d}{M_{\rm Pl}}
\end{equation}
is not only allowed but could naturally generate a $\mu$-term
\begin{equation}
\mu_\phi =
\frac{ \lambda_\mu \langle \phi \rangle_{\rm vac}^2 }{M_{\rm Pl}}
\end{equation}
of the required size \cite{mu}.
For example, for $ M = 10^{11}\,{\rm GeV} $ and
$ |\lambda_\mu | = 0.1 $ we get $ |\mu_\phi| = 400\,{\rm GeV} $.
From now on we will assume that the $\mu$-term is generated in this
way so that $ \mu_H = 0 $,
or at least $ |\mu_H| \lesssim |\mu_\phi| $,
and $ \left| \mu_\phi \right| \sim m_{\rm s} $.

Writing $ \phi = {\cal M} + \delta\phi $, where $ |{\cal M}| = M $
and $ \mu_\phi = \lambda_\mu {\cal M}^2 / M_{\rm Pl} $,
we get the relatively unsuppressed couplings
\begin{eqnarray}
{\cal L}_{\rm decay} & = &
2 \mu_\phi \tilde{H_u} \tilde{H_d} \frac{\delta\phi}{\cal M}
- 2 \left| \mu_\phi \right|^2
\left( \left| H_u \right|^2 + \left| H_d \right|^2 \right)
\frac{\delta\phi}{\cal M}
\nonumber \\ && \mbox{}
- 2 \mu_\phi \left( \overline{\lambda_tQt} H_d
+ \overline{\lambda_bQb} H_u
+ \overline{\lambda_\tau L\tau} H_u
+ A_\mu H_u H_d \right)
\frac{\delta\phi}{\cal M}
+ \mbox{c.c.}
\end{eqnarray}
where a tilde denotes the fermionic component of the superfield,
a bar denotes the hermitian conjugate,
and $A_\mu$ is the soft supersymmetry breaking parameter in
$ V_{\rm soft} = A_\mu \mu_\phi H_u H_d + \mbox{c.c.} $
and so $ |A_\mu| \sim m_{\rm s} $. 
The decay rate is then estimated to be
\begin{equation}
\Gamma \sim
\frac{ m_{\rm s}^3 }{ 10^2 M^2 }
\end{equation}
where we have roughly assumed $ m_\phi \sim |\mu_\phi| \sim |A_\mu|
\sim m_{\rm s} $.
We therefore get a reheat temperature
\begin{equation}
T_{\rm f} \sim \left( 1 \;{\rm to}\; 10\,{\rm GeV} \right)
\left( \frac{ 10^{11}\,{\rm GeV} }{ M } \right)
\left( \frac{ m_{\rm s} }{ 300\,{\rm GeV} } \right)^{\frac{3}{2}}
\end{equation}
which is sufficiently high.

We thus expect the flaton to have the following couplings to the MSSM
\footnote{As mentioned before,
$ \lambda_\mu \phi^3 H_u H_d / M_{\rm GUT}^2 $ is a
possible alternative to
$ \lambda_\mu \phi^2 H_u H_d / M_{\rm Pl} $.}
\begin{equation}
W_{\rm couplings} = \lambda_\nu L H_u \nu
+ \frac{1}{2} \lambda_\phi \phi \nu^2
+ \frac{ \lambda_\mu \phi^2 H_u H_d }{ M_{\rm Pl} }
\end{equation}

\subsection{The flaton potential and domain walls}
\label{pot}

In this section we consider the self-couplings of the flaton.
The flaton (or, in the case of a multi-component flaton, at least one
component of the flaton) should have a negative soft supersymmetry
breaking mass squared $ - m_\phi^2 |\phi|^2 $ to drive it away from
$\phi=0$ after thermal inflation.
It will also need a term to stabilise its potential at its vacuum
expectation value $ |\phi| = M \sim 10^{10} $ to
$ 10^{12}\,{\rm GeV} $.
The simplest possibility is\footnote{
$ W_{\rm vev} = \lambda_M \phi^5 / M_{\rm GUT}^2 $ would be an
alternative if $ M \gtrsim 3 \times 10^{11}\,{\rm GeV} $.
Again we assume the displayed case for simplicity.
One might even be able to use the renormalisation group running of
$m_\phi$ to stabilise the potential but then one would not
automatically get a value for $M$ in the correct range.}
\begin{equation}
\label{wvev}
W_{\rm vev} = \frac{ \lambda_M \phi^4 }{ 4 M_{\rm Pl} }
\end{equation}
and this is what we shall assume.
In the case of a multi-component flaton this could be interpreted as
for example $ \lambda_M \phi_1^4 / 4 M_{\rm Pl} $,
$ \lambda_M \phi_1 \phi_2^3 / M_{\rm Pl} $ or a sum of such terms.
See Ref.~\cite{axion} for an explicit multi-component example. 
Here, for simplicity, we will focus on the case of a single component
flaton, though it should be born in mind that a multi-component
flaton might be preferable from the model building point of view.

We then get the following scalar potential
\begin{equation}
V(\phi) = V_0 - m_\phi^2 |\phi|^2
+ \left( \frac{ A_M \lambda_M \phi^4 }{ M_{\rm Pl} }
 + \mbox{c.c.} \right)
+ \frac{ \left| \lambda_M \right|^2 \left| \phi \right|^6 }
{ M_{\rm Pl}^2 }
\end{equation}
where $ m_\phi \sim |A_M| \sim m_{\rm s} $.
This potential has four degenerate minima with $ |\phi| = M $, where
\begin{equation}
M^2 = \frac{ 2 m_\phi M_{\rm Pl} }{ 3 |\lambda_M| }
\left( \frac{|A_M|}{m_\phi}
+ \sqrt{ \frac{|A_M|^2}{m_\phi^2} + \frac{3}{4} } \right)
\end{equation}
For example, for $ m_\phi = |A_M| = 300 \,{\rm GeV} $ and
$ |\lambda_M| = 0.1 $, we get $ M = 10^{11} \,{\rm GeV} $. 
The eigenvalues of the mass squared matrix at the minima are
\begin{equation}
m_{\delta\phi_\theta}^2 = \frac{ 16 m_\phi |A_M| }{ 3 }
\left( \frac{|A_M|}{m_\phi}
+ \sqrt{ \frac{|A_M|^2}{m_\phi^2} + \frac{3}{4} } \right)
\end{equation}
and
\begin{equation}
m_{\delta\phi_r}^2 = 4 m_\phi^2 + m_{\delta\phi_\theta}^2
= 4 m_\phi^2 \left[ 1 + \frac{ 4 |A_M| }{ 3 m_\phi }
\left( \frac{|A_M|}{m_\phi}
+ \sqrt{ \frac{|A_M|^2}{m_\phi^2} + \frac{3}{4} } \right) \right]
\end{equation}
and the flatino mass squared is
\begin{equation}
m_{\tilde{\delta\phi}}^2
= 3 m_\phi^2 + \frac{3}{2} m_{\delta\phi_\theta}^2
= 4 m_\phi^2 
\left( \frac{|A_M|}{m_\phi}
+ \sqrt{ \frac{|A_M|^2}{m_\phi^2} + \frac{3}{4} } \right)^2
\end{equation}
Requiring zero cosmological constant at the minima gives
\begin{equation}
V_0 = \frac{2}{3} m_\phi^2 M^2 \left[ 1 + \frac{2|A_M|}{3m_\phi}
\left( \frac{|A_M|}{m_\phi} + \sqrt{ \frac{|A_M|^2}{m_\phi^2}
+ \frac{3}{4} } \right) \right]
\end{equation}

With four degenerate minima we clearly have to worry about a
potential domain wall problem.
The simplest way to eliminate the domain walls is to add a small term
which breaks the degeneracy of the vacua, the difference in pressure
exerted on the walls causing the domains with greater vacuum energy
to collapse.
They collapse before the domain walls come to dominate the energy
density if the difference in vacuum energies $\epsilon$ satisfies
$ \epsilon \gtrsim \sigma^2 / M_{\rm Pl}^2 $ where $\sigma$ is the
energy per unit area of the domain walls \cite{walls}.
For flaton domain walls $ \sigma \sim m_{\rm s} M^2 $ and so we
require
\begin{equation}
\epsilon \gtrsim \frac{ m_{\rm s}^2 M^4 }{ M_{\rm Pl}^2 }
\end{equation}
Therefore a term in the superpotential of the form
\begin{equation}
W_{\rm walls} \gtrsim \frac{ m_{\rm s} M^{4-n} \phi^n }
{ M_{\rm Pl}^2 }
\end{equation}
with $n$ odd would be sufficient to eliminate the domain walls.
A term with $ n = 2 \;{\rm mod}\; 4 $, for example
$ W_{\rm walls} \sim \phi^6 / M_{\rm Pl}^3 $,
would reduce the $Z_4$ domain walls to $Z_2$ domain walls.
Note that $W_{\rm walls}$ can be extremely small, and hence have a
negligible effect on the dynamics to be discussed in the next section, 
but still solve the domain wall problem.

Another way to avoid a domain wall problem is to gauge the discrete
symmetry so that there is really only one vacuum.
However, non-trivial anomaly cancellation conditions must be
satisfied \cite{dga}.
In the case of a single-component flaton with the superpotential of
Eq.~(\ref{wti}), and no extra light SU(3) multiplets,
the mixed discrete-SU(3) anomaly cancellation condition
requires $\phi^2$ to be neutral under any unbroken\footnote{
R-symmetries are broken down to $Z_2$ by hidden sector supersymmetry
breaking.}
anomaly free discrete gauge symmetry.
We therefore cannot use a discrete gauge symmetry to remove the $Z_4$
domain walls.
However, the anomaly free $Z_4$ subgroup of ${\rm U}(1)_{B-L+4Y}$,
under which $\phi$ has charge 2, can be used to gauge away the $Z_2$
domain walls left by $ W_{\rm walls} \sim \phi^6 / M_{\rm Pl}^3 $
above.
Furthermore, this symmetry is broken down to the standard matter
parity of the MSSM (which is equivalent to the R-parity of the MSSM)
by the vacuum expectation value of $\phi$.

In the case of a multi-component flaton the discrete symmetry can be
extended to a continuous symmetry which may or may not be gauged.
In the case of a continuous global symmetry, for example the
Peccei-Quinn symmetry of Ref.~\cite{axion}, the Goldstone bosons may
prove troublesome \cite{David}.
One also has more freedom to satisfy the anomaly cancellation
conditions in the case of a multi-component flaton.

\subsection{Summary}

An extension of the MSSM that gives rise to viable thermal inflation,
and so does not suffer from a moduli problem, should have the
following terms in its superpotential
\begin{equation}
\label{wti}
W_{\rm ti} = \lambda_t Q H_u t + \lambda_b Q H_d b
+ \lambda_\tau L H_d \tau + \lambda_\nu L H_u \nu
+ \frac{1}{2} \lambda_\phi \phi \nu^2
+ \frac{ \lambda_\mu \phi^2 H_u H_d }{ M_{\rm Pl} }
+ \frac{ \lambda_M \phi^4 }{ 4 M_{\rm Pl} }
\end{equation}

\section{$LH_u$ Affleck-Dine baryogenesis after thermal inflation}
\label{AD}

To orientate the reader we will first sketch the basic idea
we have in mind before plunging into the details.

The $D$-flat direction parametrised by $LH_u$ provides an ideal
Affleck-Dine field \cite{Affleck,Murayama,Dine}.
In order for it to behave as an Affleck-Dine field we must first get
it away from zero.
We therefore require its mass squared at the end of thermal
inflation, $ \left( m_L^2 - m_{H_u}^2 + |\mu_H|^2 \right) / 2 $,
to be negative.
It is simplest to assume that it rolls away from zero before $\phi$
does.
However, when $\phi=0$ the right-handed neutrinos are light, and so
$LH_u$ is not $F$-flat (it has a quartic term in its potential coming
from the superpotential).
$LH_u$ will thus be stabilised at a modest value.

Next $\phi$ will roll away from zero.
The right-handed neutrinos become heavy and so can be integrated out
leaving the effective seesaw coupling $W_{\rm seesaw}$ given in
Eq.~(\ref{wseesaw}).
This is now the term that stabilises the $LH_u$ direction and we see
that it gets smaller, and so the $LH_u$ direction
gets flatter, as $\phi$ gets larger.
Thus, as $\phi$ rolls away from zero, $LH_u$ will roll further
away from zero.
Furthermore, the soft supersymmetry breaking term derived from
$W_{\rm seesaw}$ will correlate the phases of $\phi$ and $LH_u$.

When $\phi$ becomes sufficiently large,
it will start to feel the basin of attraction of one of the minima of
its potential, and so will start curving in towards that minimum,
\mbox{i.e.} its phase will be roughly determined modulo $\pi/2$.
The phase of $LH_u$ will then be roughly determined modulo $\pi/4$ by
the soft supersymmetry breaking term derived from $W_{\rm seesaw}$.

When $|\phi|$ becomes of order $M$, a cross term from the
supersymmetric part of the potential becomes significant and changes
the correlation between the phases of $\phi$ and $LH_u$, and so gives
the phase of $LH_u$ a kick.
The direction of the kick is determined by the parameters in the
lagrangian (this is our $CP$ violation) and so gives a non-zero net
contribution when averaged over different spatial locations, unlike
the rest of the angular momentum that is flying around.
Furthermore, as this is happening $W_{\rm decay}$
(see Eq.~(\ref{wdecay})) starts to give a significant contribution
to the mass of $H_u$ and hence $LH_u$.
For $ m_L^2 - m_{H_u}^2 + |\mu_H + \mu_\phi|^2 > 0 $
this gives the $LH_u$ direction an overall positive mass squared
(as it must because $LH_u$ has a positive mass squared in the true
vacuum), and so sends $LH_u$ spiralling back in towards zero.

The effective friction on the motion of $\phi$ and $LH_u$ coming
from the Hubble expansion is negligible.
However, the effective mass squareds of both $\phi$ and $LH_u$ have
been changing sign during the above dynamics and so one would expect
them both to decay via broad parametric resonance \cite{pr}.
This will lead to approximately critical damping, and so it seems
reasonable to expect that both $\phi$ and $LH_u$ will be trapped
near their vacuum expectation values essentially immediately after
the dynamics described above has occurred.
Once they are trapped, parametric resonance becomes less efficient
because the mass squareds are now always positive.
$LH_u$'s potential near $LH_u=0$ conserves angular momentum, or in
other words lepton number, and so $LH_u$'s newly acquired lepton
number is conserved.

The dynamics outlined above is illustrated in Figure~1.

The decay of the $LH_u$ Affleck-Dine condensate will generate enough
partial reheating to restore the electroweak symmetry, and so its
lepton number can be converted to baryon number by the usual
electroweak effects \cite{Fukugita}.
Note that the energy density is still dominated by the flaton and the
reheating in the Affleck-Dine sector has a negligible effect on the
now decoupled flaton.
Finally, after the temperature has dropped to a few GeV, the flaton
decay will complete, releasing substantial entropy.

\subsection{Estimating the baryon asymmetry}

Our basic model is
\begin{equation}
W_{\rm ti} = \lambda_t Q H_u t + \lambda_b Q H_d b
+ \lambda_\tau L H_d \tau + \lambda_\nu L H_u \nu
+ \frac{1}{2} \lambda_\phi \phi \nu^2
+ \frac{ \lambda_\mu \phi^2 H_u H_d }{ M_{\rm Pl} }
+ \frac{ \lambda_M \phi^4 }{ 4 M_{\rm Pl} }
\end{equation}

The squark fields have no linear terms in their potential and have
positive mass squareds.
They will therefore be held at zero apart from thermal fluctuations,
and so can be ignored apart from their contribution to the finite
temperature effective potential.
The zero temperature potential for the other fields is
\begin{eqnarray}
V & = &
\left| \lambda_\tau L H_d \right|^2
+ \left| \lambda_\nu L H_u + \lambda_\phi \phi \nu \right|^2
+ \left| \lambda_\tau H_d \tau + \lambda_\nu H_u \nu \right|^2
+ \left| \lambda_\nu L \nu
 + \frac{ \lambda_\mu \phi^2 H_d }{ M_{\rm Pl} } \right|^2
\nonumber \\ && \mbox{}
+ \left| \lambda_\tau L \tau
 + \frac{ \lambda_\mu \phi^2 H_u }{ M_{\rm Pl} } \right|^2
+ \left| \frac{1}{2} \lambda_\phi \nu^2
 + 2 \frac{ \lambda_\mu \phi H_u H_d }{ M_{\rm Pl} }
 + \frac{ \lambda_M \phi^3 }{ M_{\rm Pl} } \right|^2
+ \mbox{$D$-terms}
\nonumber \\ && \mbox{}
+ \left( A_\tau \lambda_\tau L H_d \tau
+ A_\nu \lambda_\nu L H_u \nu
+ A_\phi \lambda_\phi \phi \nu^2
+ \frac{ A_\mu \lambda_\mu \phi^2 H_u H_d }{ M_{\rm Pl} }
+ \frac{ A_M \lambda_M \phi^4 }{ M_{\rm Pl} }
+ \mbox{c.c.} \right)
\nonumber \\ && \mbox{}
+ m_{\tau}^2 \left| \tau \right|^2
+ m_{\nu}^2 \left| \nu \right|^2
+ m_{L}^2 \left| L \right|^2
- m_{H_u}^2 \left| H_u \right|^2
+ m_{H_d}^2 \left| H_d \right|^2
- m_{\phi}^2 \left| \phi \right|^2
\end{eqnarray}
where the $m$'s and the magnitudes of the $A$'s are of order
$m_{\rm s}$.
We assume
\begin{equation}
\left. m_{LH_u}^2 \right|_{\phi=0}
= \frac{1}{2} \left( m_{L}^2 - m_{H_u}^2 \right) < 0
\end{equation}
so that the $D$-flat direction parametrised by $LH_u$ is also
unstable, in addition to the flaton $\phi$.
Note that after $\phi$ acquires its vacuum expectation value
$ M \sim \sqrt{ m_{\rm s} M_{\rm Pl} / |\lambda_M| } $,
it will give an extra contribution
$ |\lambda_\mu|^2 M^4 / M_{\rm Pl}^2 $ to $H_u$'s mass squared.
This will be of order $m_{\rm s}^2$ if
$ |\lambda_\mu| \sim |\lambda_M| $.
We assume
\begin{equation}
\left. m_{LH_u}^2 \right|_{|\phi|=M}
= \frac{1}{2} \left( m_{L}^2 - m_{H_u}^2
+ \frac{ \left| \lambda_\mu \right|^2 M^4 }{ M_{\rm Pl}^2 } \right)
> 0
\end{equation}
so that the $LH_u$ direction is stable in the true vacuum.

A rigorous study of the dynamics of this model is beyond the scope
of this paper.
Instead we will make some simplifying assumptions in order to
illustrate how the Affleck-Dine mechanism might be implemented after
thermal inflation and to crudely estimate the resultant baryon
asymmetry.

We assume that all fields are initially held at zero by the finite
temperature during thermal inflation.
We assume that the $LH_u$ direction rolls away from zero first.
It will be quickly stabilised by the term
$ | \lambda_\nu L H_u |^2 $ at a value
$ |LH_u| \sim m_{LH_u}^2 / |\lambda_\nu|^2 $.
The term $ A_\nu \lambda_\nu L H_u \nu + \mbox{c.c.} $ then causes
$\nu$ to roll away from zero.
Then the term
$ \overline{\lambda_\nu LH_u} \lambda_\phi \nu \phi + \mbox{c.c.} $
causes $\phi$ to roll away from zero\footnote{
One might imagine that our Affleck-Dine type mechanism could also be
implemented using say the right-handed electron sneutrino, which
could plausibly have a small quartic coupling $\lambda_{\phi e}$,
instead of $LH_u$.
However, unlike $LH_u$, if it was unstable it would roll away from
zero at some early time because all its couplings would be small.
The term $ A_\phi \lambda_\phi \nu^2 \phi + \mbox{c.c.} $ would then
cause $\phi$ to roll away from zero causing a premature end to
thermal inflation.}
in the direction
\begin{equation}
\label{init}
\arg \phi \simeq \arg \left[ \bar{\lambda}_\phi A_\nu
\left( \lambda_\nu L H_u \right)^2 \right]
\end{equation}
For simplicity we will assume that $\tau$ and $H_d$ remain at zero,
or at least that any expectation values they acquire can be
neglected.
With this assumption, once $\phi$ and $LH_u$ escape beyond the reach
of the temperature, their dynamics will be governed by the zero
temperature potential
\begin{eqnarray}
V & = &
\left| \lambda_\nu L H_u + \lambda_\phi \phi \nu \right|^2
+ \left| \lambda_\nu H_u \nu \right|^2
+ \left| \lambda_\nu L \nu \right|^2
+ \left| \frac{ \lambda_\mu \phi^2 H_u }{ M_{\rm Pl} } \right|^2
+ \left| \frac{1}{2} \lambda_\phi \nu^2
 + \frac{ \lambda_M \phi^3 }{ M_{\rm Pl} } \right|^2
\nonumber \\ && \mbox{}
+ \mbox{$D$-terms}
+ \left( A_\nu \lambda_\nu L H_u \nu
+ A_\phi \lambda_\phi \phi \nu^2
+ \frac{ A_M \lambda_M \phi^4 }{ M_{\rm Pl} }
+ \mbox{c.c.} \right)
\nonumber \\ && \mbox{}
+ m_{\nu}^2 \left| \nu \right|^2
+ m_{L}^2 \left| L \right|^2
- m_{H_u}^2 \left| H_u \right|^2
- m_{\phi}^2 \left| \phi \right|^2
\end{eqnarray}
As $|\phi|$ increases, $\nu$ will quickly acquire a large mass
$ \sim | \lambda_\phi \phi | $, and so will be constrained to the
minimum of its potential
\begin{equation}
\nu \simeq - \frac{ \lambda_\nu L H_u }{ \lambda_\phi \phi } 
\end{equation}
The effective potential then becomes
\begin{eqnarray}
V & = &
\left( \left| \lambda_\nu H_u \right|^2
 + \left| \lambda_\nu L \right|^2 \right)
\left| \frac{ \lambda_\nu L H_u }{ \lambda_\phi \phi } \right|^2
+ \left| \frac{ \lambda_\mu \phi^2 H_u }{ M_{\rm Pl} } \right|^2
+ \left| \frac{ \lambda_M \phi^3 }{ M_{\rm Pl} } \right|^2
+ \mbox{$D$-terms}
\nonumber \\ && \mbox{}
+ \left[ \left( A_\phi - A_\nu
 + \frac{ \bar{\lambda}_M \bar{\phi}^3 }{ 2 \phi } \right)
\frac{ \left( \lambda_\nu L H_u \right)^2 }{ \lambda_\phi \phi } 
+ \frac{ A_M \lambda_M \phi^4 }{ M_{\rm Pl} }
+ \mbox{c.c.} \right]
\nonumber \\ && \mbox{}
+ m_{L}^2 \left| L \right|^2
- m_{H_u}^2 \left| H_u \right|^2
- m_{\phi}^2 \left| \phi \right|^2
\end{eqnarray}
We assume the $D$-terms constrain $L$ and $H_u$ to the 
$D$-flat direction $LH_u$.
Then writing $ LH_u = \psi^2 / 2 $ we get
\begin{eqnarray}
V & = & \left|
\frac{ \lambda_{\nu}^2 \psi^3 }{ 2 \lambda_\phi \phi } \right|^2
+ \frac{1}{2}
 \left| \frac{ \lambda_\mu \phi^2 \psi }{ M_{\rm Pl} } \right|^2
+ \left| \frac{ \lambda_M \phi^3 }{ M_{\rm Pl} } \right|^2
- m_{\psi}^2 \left| \psi \right|^2
- m_{\phi}^2 \left| \phi \right|^2
\nonumber \\ && \mbox{}
+ \left[ \left( A_\phi - A_\nu 
 + \frac{ \bar{\lambda}_M \bar{\phi}^3 }{ 2 \phi } \right)
 \frac{ \lambda_{\nu}^2 \psi^4 }{ 4 \lambda_\phi \phi } 
+ \frac{ A_M \lambda_M \phi^4 }{ M_{\rm Pl} }
+ \mbox{c.c.} \right]
\end{eqnarray}
where $ m_{\psi}^2 = ( m_{H_u}^2 - m_{L}^2 ) / 2 $.
To make this potential more transparent, we make the following change
of variables
\begin{equation}
V = m_{\rm s}^2 M^2 \tilde{V} \,,\;
\phi = M \tilde{\phi} \,,\;
\psi = M \tilde{\psi} \,,\;
m_\phi = m_{\rm s} a_\phi \,,\;
m_\psi = m_{\rm s} a_\psi
\end{equation}
\begin{equation}
A_\nu - A_\phi = m_{\rm s} \alpha_\nu \,,\;
A_M = m_{\rm s} \alpha_M \,,\;
\lambda_\mu = \frac{m_{\rm s}M_{\rm Pl}}{M^2} \beta_\mu \,,\;
\lambda_M = \frac{m_{\rm s}M_{\rm Pl}}{M^2} \beta_M \,,\;
\frac{\lambda_\phi}{\lambda_\nu^2} = \frac{M}{m_{\rm s}} \gamma
\end{equation}
where the $a$'s and the magnitudes of the $\alpha$'s and $\beta$'s
are of order one and we assume $ |\gamma| \ll 1 $.
We then get
\begin{eqnarray}
\tilde{V} & = &
- a_\phi^2 \left|  \tilde{\phi} \right|^2
+ \left| \beta_M \right|^2 \left| \tilde{\phi} \right|^6
+ \left( - a_\psi^2
+ \left| \frac{ \tilde{\psi}^2 }{ 2 \gamma \tilde{\phi} } \right|^2
+ \frac{1}{2} \left| \beta_\mu  \right|^2
\left| \tilde{\phi} \right|^4 \right) \left| \tilde{\psi} \right|^2
\nonumber \\ && \mbox{}
+ \left[ \alpha_M \beta_M \tilde{\phi}^4
- \left( \alpha_\nu - \alpha_M
\frac{ \bar{\alpha}_M \bar{\beta}_M  \bar{\tilde{\phi}}^4 }
{ 2 \left| \alpha_M \tilde{\phi} \right|^2 } \right)
\frac{ \tilde{\psi}^4 }{ 4 \gamma \tilde{\phi} } 
+ \mbox{c.c.} \right]
\label{V}
\end{eqnarray}

When $ \tilde{\phi} \ll 1 $,
$\tilde{\psi}$'s potential is stabilised at
\begin{equation}
\label{npsi}
\left| \tilde{\psi} \right|^2
\sim \left| \gamma \right| \left| \tilde{\phi} \right|
\end{equation}
while its phase is coupled to that of $\tilde{\phi}$ by the
term\footnote{
The correlation induced by this term is different from that of
Eq.~(\ref{init}) and so the phase of $\psi$ will get a kick in the
direction $ \sin ( \arg A_\phi - \arg A_\nu ) $ while the phase of
$\phi$ will get a kick in the opposite direction.
This may contribute to the net lepton number generated,
in addition to the similar effect to be described below.}
\begin{equation}
\label{phase}
- \left( \alpha_\nu - \alpha_M
\frac{ \bar{\alpha}_M \bar{\beta}_M  \bar{\tilde{\phi}}^4 }
{ 2 \left| \alpha_M \tilde{\phi} \right|^2 } \right)
\frac{ \tilde{\psi}^4 }{ 4 \gamma \tilde{\phi} } 
+ \mbox{c.c.}
\end{equation}
the second term in the brackets being negligible at this stage.

When $ \tilde{\phi}^3 \gtrsim \gamma $, the potential for the phase
of $\tilde{\phi}$ will be dominated by the term
\begin{equation}
\alpha_M \beta_M \tilde{\phi}^4 + \mbox{c.c.}
\end{equation}
and so in some sense we can regard the phase of $\tilde{\phi}$ as
being determined modulo $\pi/2$.
Put in a different way, $\tilde{\phi}$ will be pulled towards one of
the minima of its potential and so its phase will be strongly biased
towards
\begin{equation}
\alpha_M \beta_M \tilde{\phi}^4 =
- \left| \alpha_M \beta_M \tilde{\phi}^4 \right|
\end{equation}
The phase of $\tilde{\psi}$ is then determined modulo $\pi/8$ by the
term in Eq.~(\ref{phase}).

When $\tilde{\phi}$ becomes of order one, two things happen.
First, the second term in the brackets in Eq.~(\ref{phase}) becomes
of order one and gives the phase of $\tilde{\psi}$ a kick in the
direction $ \sin ( \arg \alpha_\nu - \arg \alpha_M ) $.
Note that even before this $\tilde{\psi}$ will have had some angular
momentum about $\tilde{\psi}=0$, but it averages out to zero in the
Universe as a whole, as is shown in Figure~2(a).
This new contribution has a direction determined by the parameters of
the lagrangian and so will give a non-zero net contribution, as is
shown in Figure~2(b).
Put another way, the difference in phase between $\alpha_\nu$ and
$\alpha_M$ is our source of $CP$ violation.
Secondly, the last term in the brackets in
\begin{equation}
\left( - a_\psi^2
+ \left| \frac{ \tilde{\psi}^2 }{ 2 \gamma \tilde{\phi} } \right|^2
+ \frac{1}{2} \left| \beta_\mu  \right|^2
\left| \tilde{\phi} \right|^4 \right) \left| \tilde{\psi} \right|^2
\end{equation}
becomes of order one giving $\tilde{\psi}$ a net positive mass
squared and so causing it to spiral back in to $\tilde{\psi}=0$.

Assuming the expected broad parametric resonance \cite{pr} provides
enough damping, both $\phi$ and $\psi$ should then become trapped
near their vacuum expectation values, after which the parametric
resonance becomes less efficient.
$\psi$'s potential near $\psi=0$ conserves angular momentum, or in
other words lepton number, and so $\psi$'s lepton number is
conserved.

The Affleck-Dine condensate $\psi$ will decay well before the Hubble
expansion reduces its amplitude to the electroweak scale, and so will
release enough thermal energy to restore the electroweak symmetry.
The lepton asymmetry will then be converted into a baryon asymmetry
\begin{equation}
\frac{n_B}{s} \sim \frac{1}{3} \left( \frac{n_L}{s} \right)
\end{equation} 
by the usual electroweak effects \cite{Fukugita,Olive}.
Finally, after the temperature has dropped to a few GeV, the flaton
decay will complete, releasing substantial entropy.

The baryon asymmetry generated in this way is roughly estimated to be
\begin{eqnarray}
\frac{n_B}{s} & \sim & \frac{T_{\rm f} n_L}{m_\phi n_\phi}
\sim \frac{\theta T_{\rm f} n_\psi}{m_\phi n_\phi}
\sim \frac{ \theta |\lambda_\phi| T_{\rm f} }{ |\lambda_\nu|^2 M }
\sim \frac{ \theta ( 100\,{\rm GeV} )^2 T_{\rm f} }{ m_{\nu_L} M^2 } 
\\
& \sim & 10^{-10} \theta
\left( \frac{10\,{\rm eV}}{m_{\nu_L}} \right)
\left( \frac{T_{\rm f}}{{\rm GeV}} \right)
\left( \frac{10^{11}\,{\rm GeV}}{M} \right)^2
\label{nb}
\end{eqnarray}
where we have used Eqs.~(\ref{npsi}) and~(\ref{seesaw}).
$\theta$ is defined by this equation and will depend on the
phase difference between $\alpha_\nu$ and $\alpha_M$ as well as the
detailed dynamics.

As discussed in Section~\ref{ti}, we expect $ T_{\rm f} \sim 1 $ to
$10\,{\rm GeV} $ and $ M \sim 10^{10} $ to $10^{12}\,{\rm GeV} $.
Neutrino phenomenology \cite{mdm,solar} suggests
$ m_{\nu_{\tau L}} \sim 5\,{\rm eV} $,
$ m_{\nu_{\mu L}} \sim 10^{-2} $ to $ 10^{-3}\,{\rm eV} $ and
$ m_{\nu_{e L}} \lesssim  m_{\nu_{\mu L}} $.
Therefore, in order for Eq.~(\ref{nb}) to give the baryon asymmetry
of Eq.~(\ref{baryon}), we require $\theta$ to be roughly\footnote{
$ \theta \gtrsim 1 $ corresponds to the scenario being unviable.}
\begin{equation}
\theta_{\tau} \sim 10^{-4} \;{\rm to}\; 10
\end{equation}
\begin{equation}
\theta_e \lesssim \theta_{\mu} \sim 10^{-7} \;{\rm to}\; 10^{-2}
\end{equation}
depending on which generations make up the Affleck-Dine $LH_u$
direction.
The eventual measurement of the Higgs and slepton masses should
help to determine which of these ranges is the appropriate one
(or rule out the whole scenario), and a measurement of
$ m_{\nu_{e L}} $ would narrow the uncertainty in $\theta_e$.

\section{Conclusions}
\label{con}

Right-handed neutrinos should acquire their masses due to the vacuum
expectation value of the flaton that gives rise to thermal inflation,
not some composite GUT operator.
This will have important implications for GUT model building.
In particular, $\rm{SO}(10)$ GUT's are strongly disfavoured because
the flaton would have to be in a {\bf 126} representation which is
difficult to derive from superstrings and one would have a flaton-125
splitting problem in addition to the usual doublet-triplet splitting
problem.

The $\mu$-term of the MSSM should also be generated by the vev of the
flaton.

Our Affleck-Dine type mechanism generates a baryon asymmetry which is
roughly estimated to be
\begin{equation}
\frac{n_B}{s} \sim 10^{-10} \theta
\left( \frac{10\,{\rm eV}}{m_{\nu_L}} \right)
\left( \frac{T_{\rm f}}{{\rm GeV}} \right)
\left( \frac{10^{11}\,{\rm GeV}}{M} \right)^2
\end{equation}
where $\theta$ is the lepton asymmetry per Affleck-Dine particle.
$\theta$ depends on the difference in phase between the soft
supersymmetry breaking parameters of $W_{\rm seesaw}$ and
$W_{\rm vev}$ (\mbox{c.f.} Eqs.~(\ref{wseesaw}) and~(\ref{wvev})),
as well as the detailed dynamics. 

We also make the following prediction
\begin{equation}
m_L^2 < m_{H_u}^2
\end{equation}
modulo renormalisation effects, where $ - m_{H_u}^2 $ is the soft
supersymmetry breaking mass squared of $H_u$, and $ m_L^2 $ is the
soft supersymmetry breaking mass squared of a lepton doublet.
 
\subsection*{Acknowledgements}
EDS thanks B. de Carlos and D. H. Lyth for helpful discussions.
EDS was supported by a Royal Society Fellowship at Lancaster
University during the early stages of this work and is now supported
by a JSPS Fellowship at RESCEU.
The work of EDS is supported by Monbusho Grant-in-Aid for JSPS
Fellows No. 95209.

\newpage

\section*{Figure captions}

\subsection*{Figure 1}
Numerical simulation to illustrate the dynamics of $\phi$ and
$ LH_u = \psi^2/2 $ after thermal inflation.
The potential of Eq.~(\ref{V}) was used with the parameters
$ a_\phi = a_\psi = \alpha_\nu = \beta_M = 1 $,
$ |\alpha_M| = \beta_\mu = 2 $,
$ \gamma = 10^{-3} $ and
$ \arg \alpha_M = 3.1 $.
The initial conditions were
$ |\phi| = |\psi| = 10^{-3} $,
$ \arg \phi = 1 $ and $ \arg \psi = 0.25 $.
A friction term $ \Gamma \dot{\phi} $ with $ \Gamma = 0.75 $
was added to the equation of motion of $\phi$ to crudely simulate
the effects of parametric resonance.
A friction term was not added for $\psi$ because it would obscure
the total lepton number generated which in reality is contained in
both the homogeneous $\psi$ field and its decay products (such as
the inhomogeneous $\psi$ modes produced by parametric resonance).

\subsection*{Figure 2}
Numerical simulation to show the non-zero net lepton number
generated.
The same parameters as in Figure~1 were used except for the
following.
Motivated by Eq.~(\ref{init}), the initial phase of $\psi$
was taken to be random while the initial phase of $\phi$ was
taken to be given by $ \arg (\phi) = 4 \arg (\psi) + C $.
The lepton number produced, as measured by
$ \arg \psi (t=100) - \arg \psi (t=0) $,
is plotted against the initial phase of $\psi$.
The dotted line gives the average value.
The plots correspond to the following values of the parameters
(a) $ \arg \alpha_M = \pi $ (which is $CP$ conserving) and
$ C = 0 $, 
(b) $ \arg \alpha_M = 3.1 $ and $ C = 0 $.

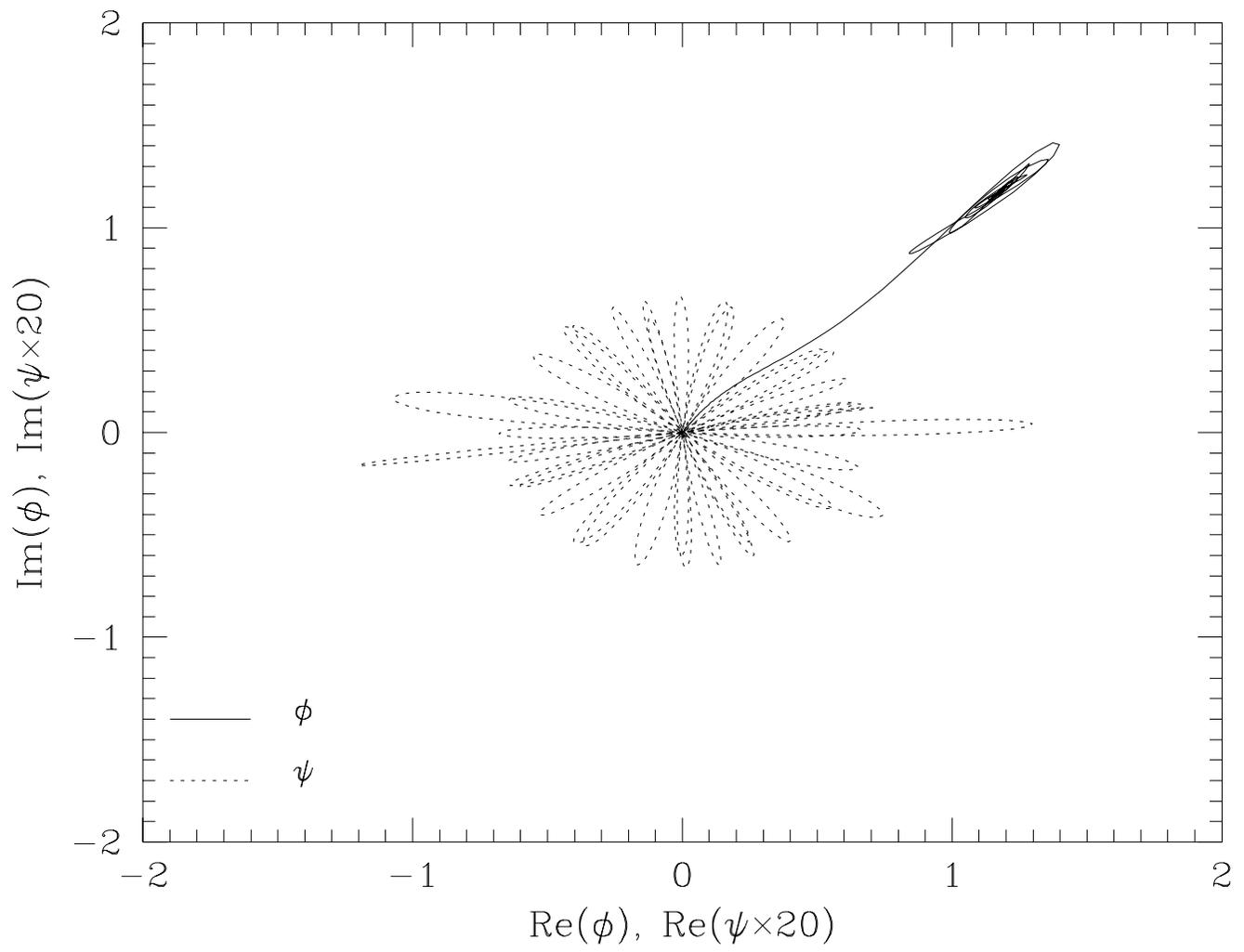
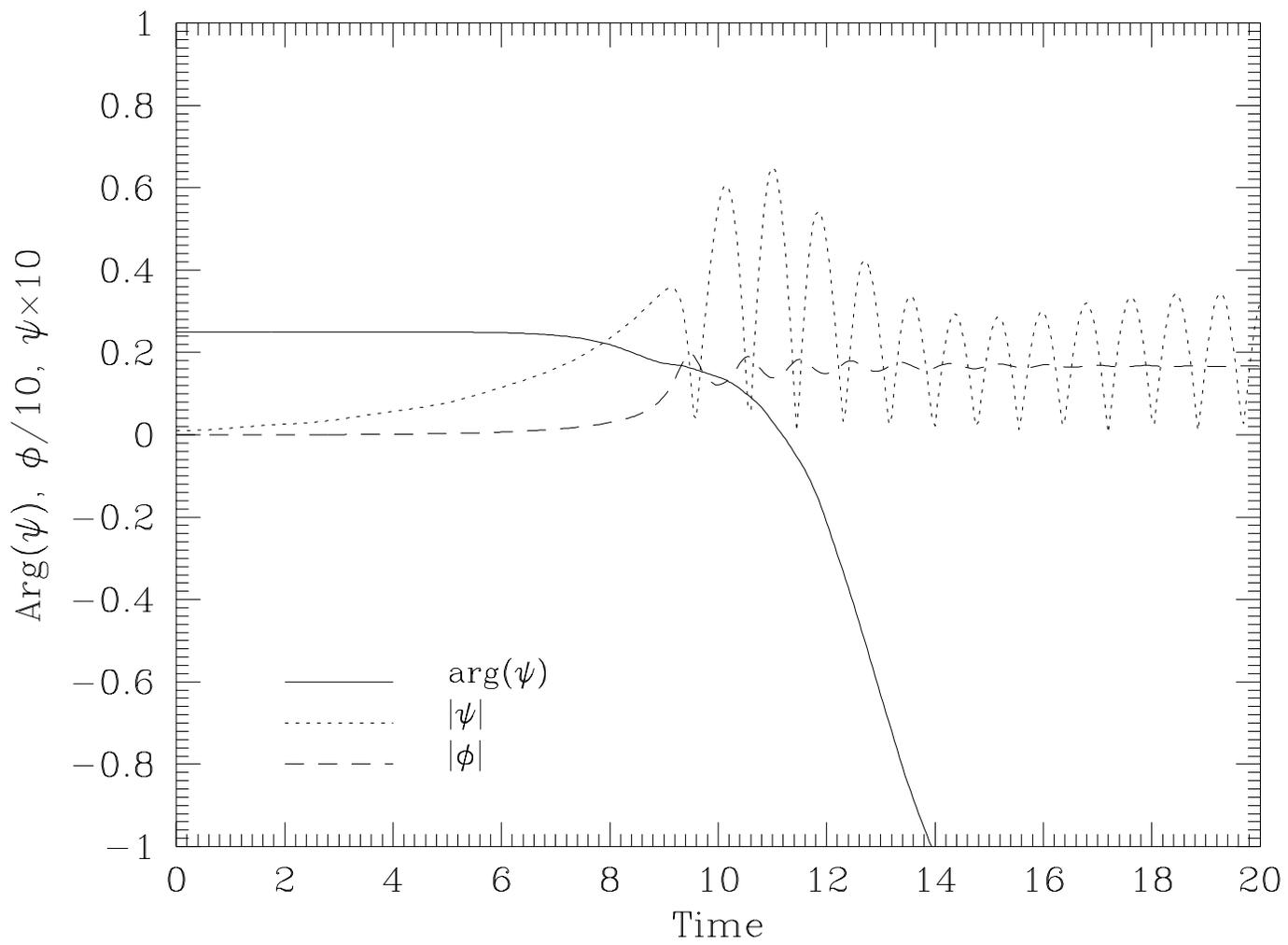
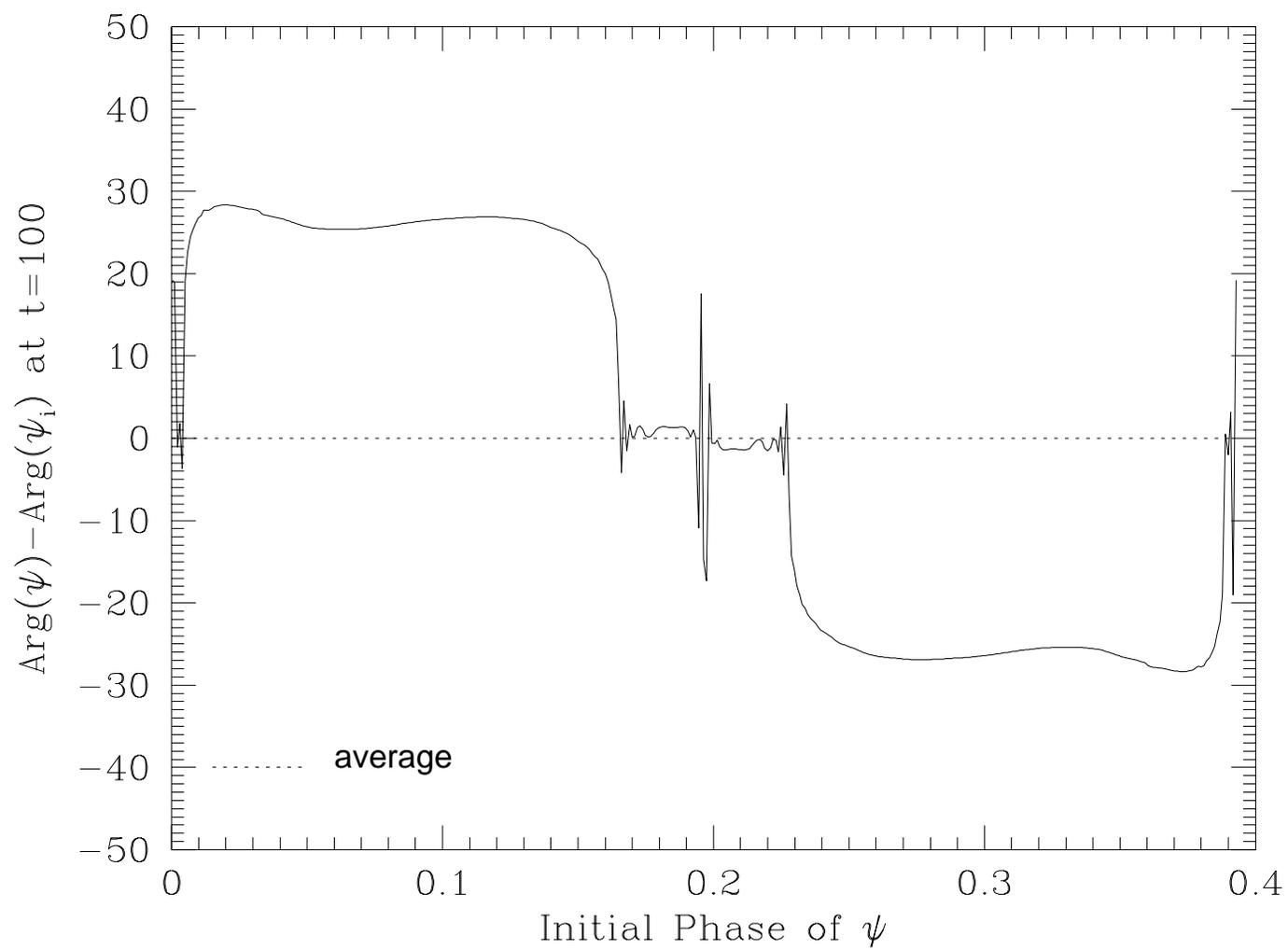
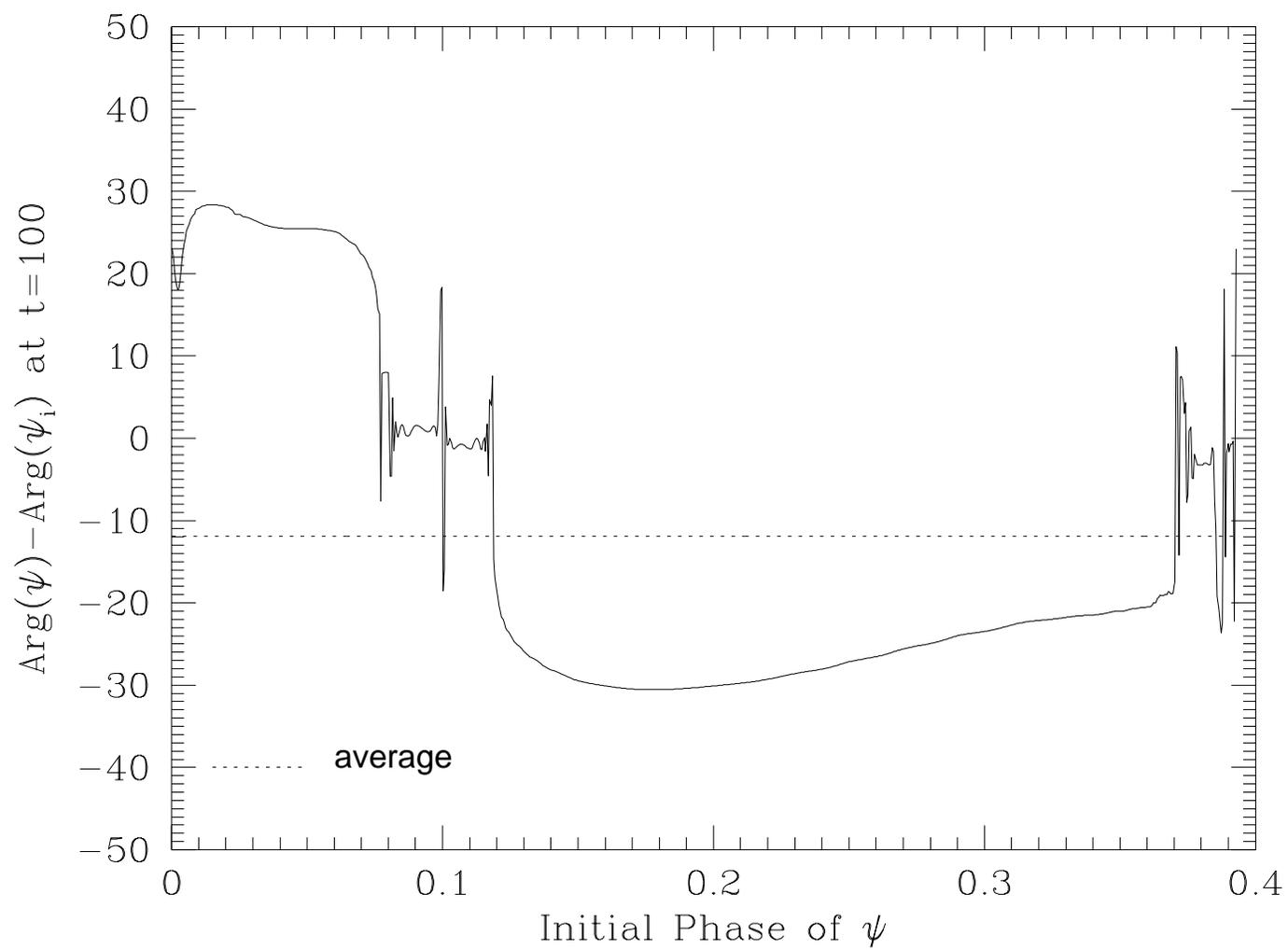
\end{document}